\documentclass[twocolumn,showpacs]{revtex4}
\usepackage{amsmath}
\usepackage{amssymb}
\usepackage{amsthm}
\usepackage{amsfonts}
\usepackage{algorithmic}
\usepackage{enumerate}
\usepackage{latexsym}
\usepackage[dvips]{graphicx}

\begin{document}

\title{Improved transfer matrix method without numerical instability}
\author{Huiqiong Yin}
\author{Ruibao Tao}
\thanks{Email: rbtao@fudan.edu.cn}
\affiliation{Department of Physics, Fudan University, Shanghai 200433, China}
\date{\today}

\begin{abstract}
A new improved transfer matrix method (TMM) is presented. It is
shown that the method not only overcomes the numerical instability
found in the original TMM, but also greatly improves the scalability
of computation. The new improved TMM has no extra cost of computing
time as the length of homogeneous scattering region becomes large.
The comparison between the scattering matrix method(SMM) and our new
TMM is given. It clearly shows that our new method is much faster
than SMM.
\end{abstract}

 \maketitle Transfer matrix
method (TMM) has been a useful approach for studying the physical
properties. There have been many publications on application of the
TMM, such as studies of Ising model \cite{Shultz,
Morgenstern[1979],Morgenstern[1980],Cheung[1983]}, quantum spin \cite%
{Suzuki[1985]}, electronic transport \cite{Sautet[1988],Wan[1998]},
and
electronic state in quasi-periodic and aperiodic chains \cite%
{Tao[1994],Huang[1998]}. TMM is also widely applied in studying the
propagation of electric-magnetic wave \cite{waveguide}, elastic wave \cite%
{stiff} and light wave \cite{Light} in multi-layer systems.

Original TMM(OTMM) can be efficiently used for the studies of
periodic system, one boundary problem between two homogeneous media
and the scattering problem of a small region sandwiched by two
leads. However, OTMM has a fundamental shortcoming due to its
numerical instability when the number of transfer steps is beyond a
critical value. This has been a major limitation to the application
of OTMM. Some approaches have been developed to calculate the
transport through a longer scattering area such as
scattering matrix method (SMM) \cite%
{Pendry[94],Pendry[other1],Pendry[other2]}, extended transfer-matrix
technique (ETMT)\cite{Wan[1998]} and Green's function approach(GFA)\cite%
{GFA}. In SMM, the middle scattering region is cut into some smaller
sections, then the scattering matrices of these sections are found
by means of TMM and combined together to get total scattering
matrix(SM) recursively. In ETMT, they used a technique to cancel the
increasing modes step by step for each slice to avoid the numerical
instability. Although the numerical instability in SMM  and ETMT
does not exist anymore, CPU time cost grows linearly when the length
of middle scattering region increases. GFA too. Based on a
continuous Schr\"{o}dinger equation in
electron wave guide without spin-orbit coupling(SOC)\cite{HuaWu} and with SOC%
\cite{YaoYang}, the stable solution of transport were also achieved.
However, there are infinite number of evanescent modes that pose
tremendous numerical difficulty, and the approximation of limit
number of evanescent wave must be done. But in some case, the
contribution of evanescent wave is significant, even more in the
presence of SOC\cite{YaoYang,LiYang}. In this letter, we present a
new improved TMM (NITMM) that has overcome such non-physical
numerical instability of OTMM. Our study will focus on the
discrete version of Schr\"{o}dinger equation with Rashba SOC\cite%
{Rashba[1960]} and will find the exact solution of electron
transport with and without SOC. Our new method not only gives
numerically stable solution at any length of the sample, but also
provides high performance in computation. That is, no extra
computing costs at increasing length of sample. Meanwhile the
simplicity of OTMM is still maintained.

\textit{General formulae of TMM.}\textbf{\ }A $2d$ strip with the
geometry of a bar sandwiched by two semi-infinite leads is studied.
The Hamiltonian of $2d$ electron gas is
\begin{equation}  \label{1}
\hat{H}=\frac{\hat{p}_{x}^{2}+\hat{p}_{y}^{2}}{2m^{\ast
}}+V(x,y)+H_{so},
\end{equation}%
where $V(x,y)$ is the potential including confinement boundary,
$H_{so}$ the remainder of the Hamiltonian that may contain
spin-orbit coupling. After discretization of the Schr\"{o}dinger
equation $\hat{H}\Psi (x,y)=E\Psi (x,y),$ the strip is replaced by a
lattice array with $N$ infinite long chains. The lattice constant is
assumed to be $a.$ Wider strip needs larger chain number $N$ to
represent$.$ The Schr\"{o}dinger equation can be
transformed into the following transfer matrix (TM) equations:%
\begin{equation}  \label{2}
\Phi _{i+1}=T_{i}\Phi _{i},\Phi
_{i+m}=(\prod\nolimits_{i=1}^{m}T_{i})\Phi _{i},
\end{equation}%
where $\Phi _{i}=(\phi _{i,N,\uparrow },\phi _{i-1,N,\uparrow },\phi
_{i,N,\downarrow },\phi _{i-1,N,\downarrow },\cdots ,\phi
_{i,1,\uparrow }$, $\phi _{i-1,1,\uparrow },\phi _{i,1,\downarrow
},\phi _{i-1,1,\downarrow })^{t}$. $\Phi _{i}$ is a matrix with
dimension $4N\times 1$. The superscript $t$ means transpose of
matrix. The element $\phi _{i,j,\sigma }$ in $\Phi _{i}$ is the
value of wave function $\Psi (x,y)$ at site $(i,j)$ with spin index
$\sigma .$ The lattice indices are $\{(i,j) \mid i\in (-\infty
,\infty ),j\in (1,N)\}$. The left lead is in the region $i\in
(-\infty ,-L-1],$ the right lead in $i\in \lbrack L+1,\infty )$, and
middle
bar in $i\in \lbrack -L+1,L-1]$. Two interfaces are at $i=-L$ and $i=L$. $%
T_{i}$ is a $4N\times 4N$ TM between the wave functions $\Phi _{i}$ and $%
\Phi _{i+1}.$ Two leads can be different or the same, and here we
assume they are the same. Further, we assume that the leads and
middle bar are homogeneous. Thus we have four different transfer
matrices: $T_{l}$ is the TM in two leads, $T_{so}$ in bar,
$T_{S_{L}}$ and $T_{S_{R}}$ at left and right interface
respectively.

Fixing adjustable parameters like hopping constant $t_{0},$ on-site
potential and the energy of electron $E$, we can obtain all elements
of four TM matrices $\{T_{l},T_{so},T_{S_{L}},T_{S_{R}}\}.$ It can
be verified that the determinants of $T_{l}$ and $T_{so}$ within
homogeneous regions are 1. Then two transform matrices $U_{l}$ and
$U_{so}$ can be numerically solved
to diagonalize $T_{l}$ and $T_{so}$: $D_{l}=U_{l}T_{l}U_{l}^{-1}$ and $%
D_{so}=U_{so}T_{so}U_{so}^{-1}$. When one knows the matrix $\Phi
_{i}$ at an arbitrary site $i,$ in principle, the wave function can
be found anywhere by TM equations. We denote $\widetilde{\Phi
}_{i}=U_{l}\Phi _{i}\ $in leads and $\widetilde{\Phi
}_{i}=U_{so}\Phi _{i}$ in middle bar. The TM equations in
diagonal representation can be written as%
\begin{equation}  \label{3}
\widetilde{\Phi }_{i+1}=\left\{
\begin{array}{l}
D_{l}\widetilde{\Phi }_{i},\ i\in (-\infty ,-L-1]\cup \lbrack
L+1,\infty ),
\\
D_{so}\widetilde{\Phi }_{i},\ i\in \lbrack -L+1,L-1].%
\end{array}%
\right.
\end{equation}%
When the strip has no interface and fully homogenous or only one
interface and two sides are homogeneous, it has been shown that the
OTMM works well and has no numerical instability. In this letter, we
study the strip with two interfaces. In this case, the OTMM will
have serious overflow problem if the length of the middle bar
between two leads is longer than a critical value.

We denote $D_{so}$ and $D_{l}$ as $D^{(c)},c=so,l.$ The diagonal elements $%
\lambda _{i}^{(c)}=D_{ii}^{(c)}$ can be classified into two types:
$|\lambda _{i}^{(c)}|=1$ and $|\lambda _{i}^{(c)}|\neq 1$ which
relate to propagating and evanescent modes respectively. For mode
$|\lambda _{i}^{(c)}|=1$, it can
be rewritten as $e^{\pm ik_{i}a}$, where $k_{i}(>0)$ is a real number, and $%
a $ is the lattice constant. $e^{ik_{i}a}$ is called as
\textquotedblleft right going\textquotedblright\ wave, and
$e^{-ik_{i}a}$ the \textquotedblleft left going\textquotedblright .
$\ $For mode $|\lambda _{i}^{(c)}|\neq 1$, it can be rewritten as
$e^{\pm (\eta _{i}a+i\phi _{i})}$
where $\eta _{i}$ is a positive real number and $\phi _{i}$ the phase. The $%
e^{\eta _{i}a+i\phi _{i}}$ is a \textquotedblleft right
growing\textquotedblright\ or say \textquotedblleft left
decaying\textquotedblright\ mode, and $e^{-\eta _{i}a-i\phi _{i}}$
the \textquotedblleft right decaying\textquotedblright\ or
\textquotedblleft left growing\textquotedblright . Due to $\det
T_{i}=\det D^{(c)}=1$ in homogenous region, any modes $e^{\pm
ik_{i}a}$ or $e^{\pm (\eta _{i}a+i\phi _{i})}$ must appear in pairs.
Then we can always arrange the diagonal
elements into the order: $diag\{D^{(c)}\}=\{e^{ik_{1}a},...,e^{ik_{p}a},e^{-%
\eta _{1}a-i\phi _{1}},\cdots ,e^{-\eta _{q}a-i\phi
_{q}},e^{-ik_{1}a},...,e^{-ik_{p}a}\newline ,e^{\eta _{1}a+i\phi
_{1}},...,e^{\eta _{q}a+i\phi _{q}}\}.$ The first $p+q$ states in
$diag\{D\}$ correspond to $p$ right going and $q$ right decaying
modes, and second $p+q$ states to $p$ left going and $q$ left
decaying modes. At eigen energy $E,$ totally we have $2p$
propagating modes, and $2q$
evanescent modes. $2p+2q=4N$, $N$ is the number of chains. For each energy $%
E,$ there are $4N$ modes distributed in $2N$ channels corresponding
to $2N$
different $|\lambda _{i}|$, where $|\lambda _{1,\cdots p}| =1 $ and $%
|\lambda _{p+1,\cdots 2N}| >1$. If $q$ equals zero, all states are
extended. When $p=0,$ no propagating wave can exist in strip.
Changing energy $E$ results in changing of the $p$ and $q.$

We assume that a right going electron wave, $e^{ik_{1}x_{i}},$ is
injected from the first channel of left lead. In general there
should be some reflection waves in all channels in left lead, due to
the scattering of
interfaces. We denote the reflection waves in $2N$ channels as $%
\{r_{l}e^{-ik_{l}x_{i}},r_{p+m}e^{\eta _{m}x_{i}+i\phi
_{m}}:l=1,2,...,p;m=1,2,...,q\}.$ $\{r_{l,}r_{p+m}\}$ describe the
reflection coefficients. The wave function in the left lead is $\widetilde{%
\Phi }%
_{i}=(e^{ik_{1}x_{i}},0,...,0,r_{1}e^{-ik_{1}x_{i}},...,r_{p}e^{-ik_{p}x_{i}},r_{p+1}e^{\eta _{1}x_{i}+i\phi _{1}},...,%
\newline
r_{p+q}e^{\eta _{q}x_{i}+i\phi _{q}})^{t}$. We can further set the
wave
function at position $i=-L$ to be $\widetilde{\Phi }%
_{-L}=(1,0,...,0,r_{1},r_{2},..,r_{2N})^{t}$. The phases of
reflection waves
have been absorbed in coefficients $\{r_{i}\}$. We have $\widetilde{\Phi }%
_{-L}=D_{l}\widetilde{\Phi }_{-L-1}$, $\widetilde{\Phi }_{-L+1}=\widetilde{T}%
_{S_{L}}\widetilde{\Phi }_{-L}$, $\widetilde{\Phi }_{L}=(D_{so})^{2L-1}%
\widetilde{T}_{S_{L}}\widetilde{\Phi }_{-L}$, $\widetilde{\Phi }_{L+1}=%
\widetilde{S}\widetilde{\Phi }_{-L},$where%
\begin{equation}  \label{4}
\left\{
\begin{array}{l}
\widetilde{T}_{S_{L}}=U_{so}T_{S_{L}}U_{l}^{-1},\widetilde{T}%
_{S_{R}}=U_{l}T_{S_{R}}U_{so}^{-1} \\
\widetilde{S}=\widetilde{T}_{S_{R}}(D_{so})^{2L-1}\widetilde{T}_{S_{L}}%
\end{array}%
\right.
\end{equation}%
Then we have $(\widetilde{\Phi }_{L+1})_{\alpha }=\sum_{\beta =1}^{4N}%
\widetilde{S}_{\alpha ,\beta }(\widetilde{\Phi }_{-L})_{\beta }=\widetilde{S}%
_{\alpha ,1}+\sum_{i=1}^{2N}\widetilde{S}_{\alpha ,2N+i}r_{i}$, where $%
\alpha =1,...,4N$. There are no left going waves injected from right
lead,
so the wave function at $i=L+1$ can be expressed by $\widetilde{\Phi }%
_{L+1}=(t_{1},t_{2},...,t_{2N},0,..,0)^{t}$, where $\{t_{i}\}$ are
the transmission coefficients. Thus we obtain $2N$ equations:
\begin{equation}  \label{5}
(\widetilde{\Phi }_{L+1})_{\alpha }=0,\alpha =2N+1,\cdots ,4N.
\end{equation}

Define $2N\times 2N$ matrices $W$ and $Y:$ $W_{lm}=\widetilde{S}%
_{2N+l,2N+m};Y_{l}=-\widetilde{S}_{2N+l,1};l,m=1,2,...,2N$ and the
reflection coefficient matrix $R=(r_{1},r_{2},...,r_{2N})^{t}.$ Equation (%
\ref{5}) can be rewritten as $WR=Y$. The matrix $R$ can be uniquely
obtained through $R=W^{-1}Y.$ Then $2N$ transmission coefficients
can be obtained from equations $t_{\alpha }=(\widetilde{\Phi
}_{L+1})_{\alpha }$, $\alpha =1,2,...,2N.$ Hence, the wave function
at every site in leads and middle bar is obtained. So basically the
electron transport though two interfaces and middle bar is
completely solved. However, the computation of $W^{-1}$ will
encounter numerical overflow problem when $L$ is large. The problem
comes from the term $(D_{so})^{2L-1}$ that may have some increasing
modes with diagonal elements like $|\lambda _{p+i}|^{2L-1}.$ If the
value of $|\lambda
_{p+i}|^{2L-1}$ $\gg1,$ many elements $W_{lm}=\sum_{i=1}^{4N}(\widetilde{T}%
_{S_{R}})_{2N+l,i}\lambda
_{ii}{}^{2L-1}(\widetilde{T}_{S_{L}})_{i,2N+m}$
are of the order of $|\lambda _{p+i}|^{2L-1}.$ Hence the calculation of $%
W^{-1}$ will meet the numerical overflow with the order of $\left(
\lambda _{p+i}{}^{2L-1}\right) ^{n},1\ll n\leq 2N$. Therefore, the
OTMM can only accurately solve the solution for bar system with
smaller length $L$. For example, we have calculated a Rashba SOC
system with $N=200$ by means of OTMM. The longest length of bar is
around $15$, and numerical instability happens for $2L>15.$ Larger
$N$ results more shorter $2L$.

\textit{New improved TM method} In fact, the physics here must be
finite. The solution for $\{r_{i}\}$ should be stable. The numerical
overflow is an artifact of the OTMM. Even $|\lambda _{i}|^{2L-1}$
$\gg1$ for evanescent modes when $2L-1\gg1,$ the wave function after
$2L-1$ steps of the transfer by $\left( D_{so}\right) ^{2L-1}$ still
should be finite. Thus the values of
$(\widetilde{T}_{S_{L}}\widetilde{\Phi
}_{-L})_{2N+p+i},i=1,2,...,q,$ corresponding to the increasing modes
of $\{|\lambda _{p+i}|>1\}$ must be
very small to assure $(\left( D_{so}\right) ^{2L-1}\widetilde{T}_{S_{L}}%
\widetilde{\Phi }_{-L})_{2N+p+i}$ to be finite. That is the physical
requirement. Thus we introduce $q$ new auxiliary parameters $\{\zeta
_{i}\}$ and assume that
\begin{equation}
(\widetilde{T}_{S_{L}}\widetilde{\Phi }_{-L})_{2N+p+i}=\zeta
_{i}e^{-(2L-1)\left( \eta _{i}a+i\phi _{i}\right) },i=1,...,q.
\label{7}
\end{equation}
$q$ parameters $\{\zeta _{i}\}$ have\ to be determined together with
$2N$ reflection coefficients $\{r_{i}\}$. The elements in matrix
$\widetilde{\Phi }_{L+1}$ is the linear combination of $2N$
coefficients $\{r_{i}\}$ and $q$
auxiliary parameters $\{\zeta _{i}\}.$ We have%
\begin{equation}  \label{8}
(\widetilde{\Phi }_{L+1})_{\alpha }=D_{\alpha
}+\sum\nolimits_{i=1}^{2N}C_{\alpha
i}r_{i}+\sum\nolimits_{j=1}^{q}B_{\alpha j}\zeta _{j}
\end{equation}%
where%
\begin{equation}
\left\{
\begin{array}{l}
D_{\alpha }=\sum\nolimits_{\beta
=1}^{2N+p}(\widetilde{T}_{S_{R}})_{\alpha
\beta }((D_{so})^{2L-1}\widetilde{T}_{S_{L}})_{\beta 1}, \\
C_{\alpha i}=\sum\nolimits_{\beta
=1}^{2N+p}(\widetilde{T}_{S_{R}})_{\alpha
\beta }((D_{so})^{2L-1}\widetilde{T}_{S_{L}})_{\beta ,2N+i}, \\
B_{\alpha i}=(\widetilde{T}_{S_{R}})_{\alpha ,2N+p+i},\ \alpha =1,2,...,4N.%
\end{array}%
\right.
\end{equation}

No left going wave in right lead is requested, i.e. $(\widetilde{\Phi }%
_{L+1})_{\alpha }=0,\alpha =2N+1,\cdots ,4N$, which yield following
$2N$ equations
\begin{equation}  \label{9}
D_{\alpha }+\sum\nolimits_{i=1}^{2N}C_{\alpha
i}r_{i}+\sum\nolimits_{j=1}^{q}B_{\alpha j}\zeta _{j}=0.
\end{equation}

$2N$ equations ({\ref{9}}) together with $q$ auxiliary equations
(\ref{7}), totally we have $2N+q$ equations that can uniquely
determine $2N+q$ unknown parameters $\{r_{i}\}$ and $\{\zeta
_{j}\}$. All coefficients of $\{\zeta _{i}\}$ and $\{r_{i}\}$ are
finite here so that the inversion of coefficient matrix can be
calculated without overflow and unique solution of $\{\zeta _{i}\}$
and $\{r_{i}\}$ are obtained. Now the numerical overflow problem
does not exist any more. The method is stable for any length. After all $%
\{r_{i},\zeta _{j}\}$ are found, the transmission coefficients
$t_{\alpha }$
can be obtained by equations%
\begin{equation}  \label{10}
t_{\alpha }=(\widetilde{\Phi }_{L+1})_{\alpha },\alpha =1,2,...,2N.
\end{equation}

As an example, we use our NITMM to study electron transport through
a Rashba bar sandwiched by two semi-infinite metal leads. The
Hamiltonian in bar region is
\begin{equation}  \label{11}
\hat{H}=\frac{\hat{p}_{x}^{2}+\hat{p}_{y}^{2}}{2m^{\ast }}+\frac{\alpha }{%
\hbar }\left(
\hat{p}_{y}\hat{\sigma}_{x}-\hat{p}_{x}\hat{\sigma}_{y}\right)
+V_{\mathrm{conf}}(x,y),
\end{equation}%
where $(\hat{\sigma}_{x},\hat{\sigma}_{y},\hat{\sigma}_{z})$ are
Pauli
matrices, $\alpha $ the strength of the Rashba SOC, and $V_{\mathrm{conf}%
}(x,y)$ the transverse confining potential. Here an open boundary
condition in $y$ direction is applied. TM in leads and Rashba bar
can be written as
following super-matrix%
\begin{equation}
T_{i}=\left(
\begin{array}{ccccc}
A & B & \cdots & 0 & 0 \\
B^{\ast } & A & \cdots & 0 & 0 \\
\vdots & \vdots & \ddots & \vdots & \vdots \\
0 & 0 & \cdots & A & B \\
0 & 0 & \cdots & B^{\ast } & A%
\end{array}%
\right),
\end{equation}%

\begin{figure}[t]
\includegraphics[width=8.5cm]{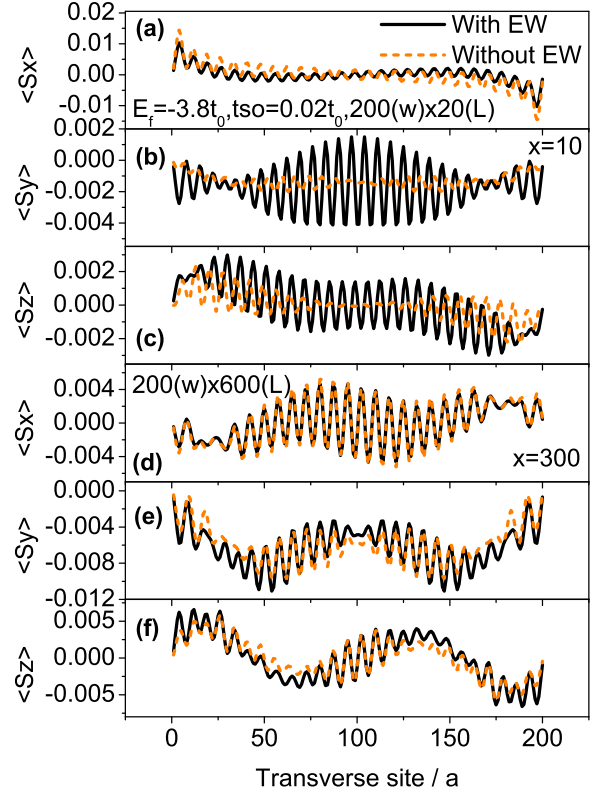}
\caption{a), b), c) are the spin polarization(x=10) for 200$\times
$20, while d), e), f)(x=300) for 200$\times $600, with the unit
$\hbar /2$. The continuous line included the evanescent waves(EW).
The dotted line included only the extending states. }
\label{fig:polarization1}
\end{figure}
where $A$ and $B$ are two $4\times 4$ sub-matrix, $A^{\ast}(B^{\ast
})$ is the complex conjugate matrix of $A$ $(B)$.
\begin{equation}
A=\left(
\begin{array}{cccc}
a & b & e & f \\
1 & 0 & 0 & 0 \\
-e & -f & a & b \\
0 & 0 & 1 & 0%
\end{array}%
\right) , B=\left(
\begin{array}{cccc}
g & 0 & h & 0 \\
0 & 0 & 0 & 0 \\
-h^{\ast } & 0 & g^{\ast } & 0 \\
0 & 0 & 0 & 0%
\end{array}%
\right)
\end{equation}%
where $\left\{ a,b,e,f,g,h\right\} $ are the functions of the
hopping constant $t_{0}$, SOC strength ($t_{so}$ in Rashba region,
$0$ in lead region) and eigen energy $E$. The expressions of
$\left\{ a,b,e,f,g,h\right\} $ are $a=-c\left( E-w_{x,y}\right)
/t_{0},b=c\left( t_{so}^{2}/t_{0}^{2}-1\right) $, $e=-ct_{so}\left(
E-w_{x,y}\right) /t_{0}^{2},$ $f=-2ct_{so}$ $/t_{0}$, $g=c\left(
it_{so}^{2}/t_{0}^{2}-1\right) $, $h=-c\left( 1-i\right) t_{so}/t_{0},$and $%
c=\left( 1+t_{so}^{2}/t_{0}^{2}\right) ^{-1}$, where $t_{0}=\hbar
^{2}/2m^{\ast }a^{2}$, $t_{so}=\alpha /2a$. $w_{x,y}$ is the on-site
energy, here we have chosen it to be zero.

\begin{figure}[t]
\includegraphics[width=8.5cm]{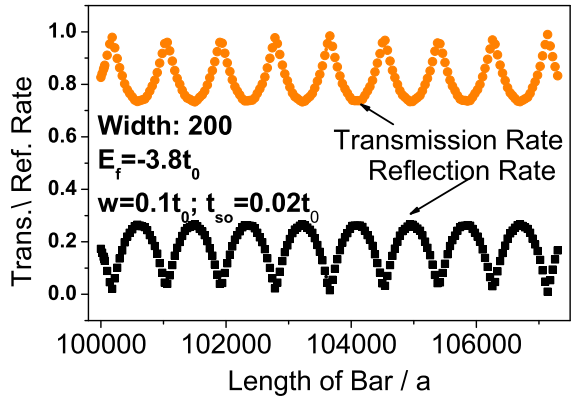}
\caption{The upper orange line stands for transmission rate, while
the lower black line is the reflection rate.}
\label{fig:oscillation}
\end{figure}
We take the $E$ to be Fermi energy $E_{F},$ and fix the other
parameters in TM, then consider an electron wave injecting from one
of the channels in
left lead and calculate the reflection and transmission coefficients $%
\{r_{i},t_{i}\}$. Finally the wave functions at every site of the
strip can be obtained. We compare the result of our new improved
method with that of the OTMM for $2L\leq 15$ for $N=200$ system
within which the calculation of OTMM is accurate and has no
numerical instability. We obtain the exact same results for the
transmission and reflection coefficients. However, our new method
can accurately calculate the case of $2L$ as large as we want. Main
cost of computer time is from the diagonalization of $800\times 800$
TM. It is clearly shown that our new method does not require more
computer time when the length of middle bar increases. Figure
~\ref{fig:polarization1} and ~\ref{fig:oscillation} give the exact
solution of the spin polarization in middle bar and resonant
transmission of current as a function of bar length up to $10^{5}$
lattices, which is a size that can not be calculated by other
methods. Where we have taken the average for non-polarized
injection. The contribution from evanescent waves to spin
polarization of Rashba SOC system based on continuous version has
been studied by Li and Yang\cite{LiYang}. From our results of
figure~\ref{fig:polarization1}, the exact calculation shows that the
contribution from evanescent wave is significant for a short bar,
but it is small when the length of bar becomes long.

\begin{figure}[h]
\includegraphics[width=8.5cm]{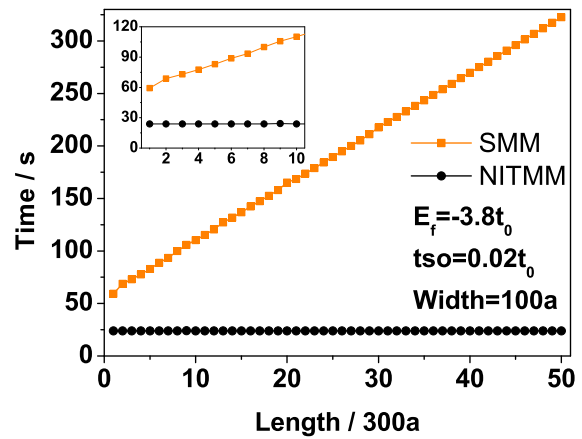}
\caption{The time cost of SMM(orange upper) increases as the first
order of lengths of scattering region, where as that of NITMM(black
below) is the zeroth. The maximum length of a single block within
which the SMM is applicable is less than 400a, for that there is a
matrix inversion operation in the SMM formulae ~\protect\cite{SMM}.
Here we choose the maximum length as 300a to ensure the precision of
calculation.} \label{fig:timecost}
\end{figure}

\textit{Comparison between our new method and previous methods (SMM
and ETMT)}. SMM has been mostly applied for studying the wave
transport and has no problem of numerical
instabilty\cite{Pendry[94]}. In SMM, in order to avoid the numerical
instability, one has to divide the middle bar into many small
sections. Then one should find the SM for each section via TMM. In
the
sample with $N=100,$ the longest length of such pieces should be less than $%
400$ lattices such that no numerical instability happens. Here we
use the 300 to ensure the precision of calculation. Then a recursion
approach is
applied to combine every sub SM to reach the final total SM\cite{SMM}. Fig.~%
\ref{fig:timecost} shows that the CPU cost of SMM linearly increases
with growing length of scattering region, but our new TMM is
independent of the length of middle bar as showed in same figure.
The numerical solutions obtained by SMM and our new TMM are exactly
the same. In addition, when one studies the transmission of a
polarized incident wave going through the middle bar, considering
only a single channel injection of electron wave is not sufficient
for SMM. All channels injection from left and right must be
calculated individually to obtain the SM. Thus, the computing time
to obtain the SM of first block near interface is almost 4N times of
NITMM. Further more, computing time grows linearly for adding every
block by recursive algorithm in SMM. In ETMT\cite{Wan[1998]}, they
studied the magnetoconductance of a quantum wire with several
antidots. The numerical technique used there could avoid numerical
instability happened in original transfer-matrix method, and was
applied to a variety of 3D systems involving complicated atomic and
many-body potentials. However, due to its iterative calculations,
the computing time of ETMT also increases linearly with the increase
in the length of quantum wire even for the homogeneous region. Our
method shows the superiority in treating long homogeneous transfer
region for that it does not cost extra computing time as the length
of homogeneous scattering region becomes large. The computing time
is in the zeroth order of homogenous region length L, $O(L^{0}).$
For the model we calculated here, our method is much faster than SMM
for long length system, and we believe that it is also faster than
ETMT.
Our method can also be applied to the studies on the transport under
uniform magnetic field. Extensions of our NITMM to other problems
will be our future work.

\acknowledgments This work is supported by the National Natural
Science Foundation of China (Nos. 10674027 and 10547001) and 973
project of China under grant No.2006CB921300.




\begin{thebibliography}{99}
\bibitem{SMM} The detail formulae of recursive SMM can be found on page 223,
formulae (56), Ref.13
\end{thebibliography}
\end{document}